# Theory of Enhanced Interlayer Tunneling in Optically Driven High $T_c$ Superconductors

Jun-ichi Okamoto,[1, 2, *] Andrea Cavalleri,[3, 4] and Ludwig Mathey[1, 2]

[1]*Zentrum für Optische Quantentechnologien und Institut für Laserphysik, Universität Hamburg, 22761 Hamburg, Germany*
[2]*The Hamburg Centre for Ultrafast Imaging, Luruper Chaussee 149, 22761 Hamburg, Germany*
[3]*Max Planck Institute for the Structure and Dynamics of Matter, 22761 Hamburg, Germany*
[4]*Department of Physics, Clarendon Laboratory, University of Oxford, Oxford OX1 3PU, United Kingdom*

Motivated by recent pump-probe experiments indicating enhanced coherent *c*-axis transport in underdoped YBCO, we study Josephson junctions periodically driven by optical pulses. We propose a mechanism for this observation by demonstrating that a parametrically driven Josephson junction shows an enhanced imaginary part of the low-frequency conductivity when the driving frequency is above the plasma frequency, implying an effectively enhanced Josephson coupling. We generalize this analysis to a bilayer system of Josephson junctions modeling YBCO. Again, the Josephson coupling is enhanced when the pump frequency is blue detuned to either of the two plasma frequencies of the material. We show that the emergent driven state is a genuine, nonequilibrium superconducting state, in which equilibrium relations between the Josephson coupling, current fluctuations, and the critical current no longer hold.

PACS numbers:

Recent pump-probe experiments on high-temperature superconductors such as YBCO revealed transiently enhanced superconducting-like states both below and above their critical temperatures $T_c$ [1–4]. The origin of these transient superconducting states has not been identified yet. Several ideas have been proposed: nonlinear phononic effects [5], parametric cooling [6, 7], competing orders [8, 9], and redistribution of phase fluctuations [10]. In this Letter, we model the layered structure of YBCO as a Josephson junction chain, and demonstrate that below the critical temperature the parametric driving of these junctions enhances the Josephson coupling in the steady state [31]. We extract this quantity from the $1/\omega$ divergence of the imaginary part of the conductivity $\sigma(\omega)$ as the frequency $\omega$ approaches zero.

So far, most of the theoretical studies have considered quantities such as the power spectrum of currents [10], or the dc current response [6]. While these quantities give important insight into the system, here we discuss the optical conductivity itself by including the probing field in our calculation to obtain the actual nonequilibrium response. The Kubo formula or other equilibrium methods are not used. This achievement is crucially important because the conclusions of Refs. 1–4 are based precisely on this quantity. We note that while experiments indicate stiffening of the superfluid density both below and above $T_c$, here we focus on cases below $T_c$, in which condensed Cooper pairs are safely assumed. We will discuss cases with fluctuating Cooper pairs just above $T_c$ (below the pseudogap temperature) elsewhere. Complete understanding how the interlayer coherence is dynamically enhanced may provide new theoretical and experimental means to investigate fluctuating orders above $T_c$.

In this Letter, we first study a single Josephson junc-tion. We show that a driving frequency just above the plasma frequency leads to strong enhancement of the Josephson coupling, derived from the low-frequency conductivity. This arises because the nonlinear driving term couples the probe pulse to the plasma frequency. This parametric mechanism can also be considered as a Fano-Feshbach resonance [11]. Then we turn to a bilayer system of Josephson junctions (Fig. 1), which has been used to explain the optical properties of YBCO [12–14]. We employ both an analytical approach and Langevin simulations to calculate the conductivities. We find that, when the driving frequency is just above the low and high plasma frequencies, the Josephson coupling increases compared to the undriven cases by 20% and 8% respectively, in the steady state (Fig. 1). We propose ways to further increase the enhancement; near the low plasma frequency, increasing capacitive couplings of both junctions is ideal, while near the high plasma frequency, increasing (decreasing) the coupling of weak (strong) junctions is preferable. Finally, we demonstrate the counterintuitive properties of this nonequilibrium superconductor. We discuss a scenario where both current fluctuations and the Josephson coupling are enhanced, which would be unexpected in equilibrium. We explain this phenomena by the redistribution of current fluctuations in frequency space.

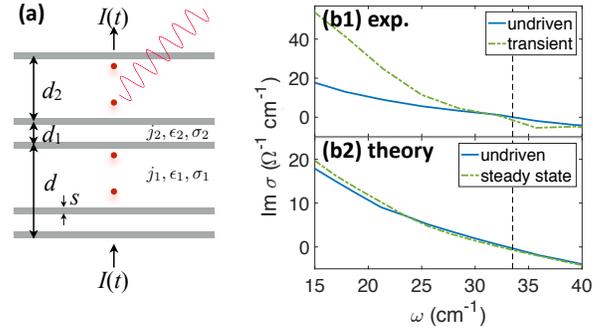

FIG. 1: (a) Schematic depiction of YBCO. Superconducting $CuO_2$ layers (grey) form series of bilayer Josephson junctions. THz pulses (wavy lines) excite apical oxigen atoms (circles) that induce oscillations of $j_1$ and $j_2$. (b) Comparison of the calculated imaginary part of the conductivity with experiments in Ref. 2 at 10K (The lower plasma frequency is $\sim 33$cm$^{-1}$.) The optimal driven conditions near the higher plasma frequency are used for the calculations. (Theoretical curves are rescaled so that undriven curves from the experiment and theory coincide.)

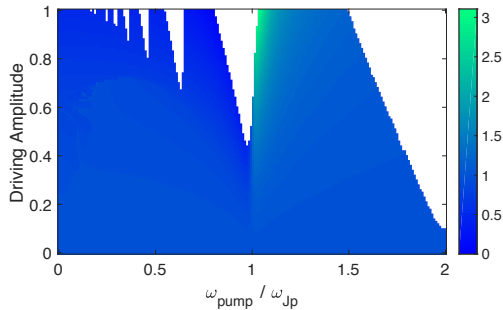

FIG. 2: The normalized effective Josephson coupling $J_{\mathrm{eff}}/J_0$. We note that a significant increase of the coupling is achieved when the driving frequency is just above the plasma frequency. Dynamically unstable regions are left blank.

We start with a single parametrically driven junction [15–20], with a bare Josephson coupling $J_0$, a thickness $d$, and a dielectric constant $\epsilon$. It has a characteristic plasma frequency $\omega_{\mathrm{Jp}} = \sqrt{4\pi e^* dJ_0/\hbar\epsilon}$. The phase $\varphi$ of the junction obeys

$$\ddot{\varphi} + \gamma\dot{\varphi} + \omega_{\mathrm{Jp}}^2 \left[1 + A\cos(\omega_{\mathrm{pump}}t)\right]\sin\varphi = \tilde{I}, \quad (1)$$

where $\gamma$ is a damping coefficient, $\tilde{I} \equiv \omega_{\mathrm{Jp}}^2 I/J_0$, and $I$ is an external current. We consider a periodic modulation of $J_0$ with an amplitude $A$ ($0 < A < 1$) at a frequency $\omega_{\mathrm{pump}}$ as $J_0 \to J_0\left[1 + A\cos(\omega_{\mathrm{pump}}t)\right]$. This induces parametric driving. $\tilde{I}$ is independent of $A$ since $\omega_{\mathrm{Jp}}^2 \propto J_0$. Linearizing and Fourier transforming the equation, we obtain

$$\left(\frac{-\omega^2 - i\gamma\omega}{\omega_{\mathrm{Jp}}^2} + 1\right)\varphi(\omega)$$
$$= -\frac{A\left[\varphi(\omega + \omega_{\mathrm{pump}}) + \varphi(\omega - \omega_{\mathrm{pump}})\right]}{2} + \tilde{I}(\omega). \quad (2)$$

To calculate the conductivity, we assume that the probing current is monochromatic $\tilde{I}(\omega) = \tilde{I}_{\mathrm{probe}}\delta(\omega - \omega_{\mathrm{probe}})$. We need to solve a discrete set of equations for $\varphi_n \equiv \varphi(\omega_{\mathrm{probe}} + n\omega_{\mathrm{pump}})$, $n \in \mathbb{Z}$, to determine $\varphi(\omega_{\mathrm{probe}})$. Since higher harmonics are of higher order in the driving amplitude, we consider only $\varphi_{\pm 1}$ and $\varphi_0$ [32]. Using the Josephson relation $V = (\hbar/e^*)\dot{\varphi}$, and $\sigma = Id/V$, we find [33]

$$\sigma(\omega_{\mathrm{probe}}) = \frac{\epsilon}{4\pi i\omega_{\mathrm{probe}}}\left[\frac{A^2\omega_{\mathrm{Jp}}^4}{4}\left(\frac{1}{\omega_{\mathrm{Jp}}^2 - (\omega_{\mathrm{pump}} - \omega_{\mathrm{probe}})(-i\gamma + \omega_{\mathrm{pump}} - \omega_{\mathrm{probe}})}\right.\right.$$
$$\left.\left. + \frac{1}{\omega_{\mathrm{Jp}}^2 - (\omega_{\mathrm{pump}} + \omega_{\mathrm{probe}})(i\gamma + \omega_{\mathrm{pump}} + \omega_{\mathrm{probe}})}\right) + \omega_{\mathrm{probe}}^2 + i\gamma\omega_{\mathrm{probe}} - \omega_{\mathrm{Jp}}^2\right]. \quad (3)$$

In the London and Ginzburg-Landau picture [21],

$$\mathrm{Im}\left[\sigma_{\mathrm{London}}(\omega)\right] = \frac{n_S e^{*2}}{m\omega} = \frac{c^2}{4\pi\lambda_L^2\omega} = \frac{J_0 e^* d}{\hbar\omega}, \quad (4)$$

where $n_S$ is the superfluid density, $m$ the mass of a Cooper pair, $\lambda_L$ the London penetration depth, and $J_0$ the bare Josephson coupling. Thus we define an effective Josephson coupling $J_{\mathrm{eff}}$ as

$$J_{\mathrm{eff}} \equiv \frac{\hbar}{e^* d}\mathrm{Im}[\sigma(\omega)\omega]_{\omega=0}$$
$$= J_0\left[1 - \frac{A^2\omega_{\mathrm{Jp}}^2(\omega_{\mathrm{Jp}}^2 - \omega_{\mathrm{pump}}^2)}{2(\omega_{\mathrm{Jp}}^2 - \omega_{\mathrm{pump}}^2)^2 + 2\gamma^2\omega_{\mathrm{pump}}^2}\right]. \quad (5)$$

The driving term generates a resonant renormalization of $J_0$, which is the key mechanism that we propose here. When the driving frequency is above (below) the plasma frequency, the effective coupling is enhanced (decreased). The maximal value of the correction $J_{\mathrm{eff}}/J_0 = 1 + A^2\omega_{\mathrm{Jp}}^2/2\gamma(\gamma + 2\omega_{\mathrm{Jp}})$ is achieved at $\omega_{\mathrm{pump}} = \omega_{\mathrm{Jp}}\sqrt{1 + \gamma/\omega_{\mathrm{Jp}}}$. We note that Eq. (5) is of the form of a Fano-Feshbach resonance [11]; the plasma mode corresponds to the bound state, and the pump-pulse couples the probing mode to this bound state.

In the Supplemental Material, we include the next order of the nonlinearity of $\sin\varphi$ in Eq. (1). We show the result of this extended analysis in Fig. 2 for $\gamma = 0.05$. We exclude the regions of dynamical instability determined by Floquet analysis of the linear model[22, 23] [32]. As the damping $\gamma$ increases, the instability regions diminish while the enhancement of $J_{\text{eff}}$ also decreases; for the optimal gain of $J_{\text{eff}}$, it is desirable to control the damping to balance these two effects.

Next we consider a bilayer system of $2N$ Josephson junctions composed of $2N+1$ superconducting (SC) layers stacked along the $c$ axis that models the electromagnetic response of YBCO [2, 13, 14] (Fig. 1). The $m$th superconducting layer has the excess charge $Q_m$ and the phase of the superconducting order parameter $\theta_m$. These are conjugate variables, satisfying $\{\theta_l, -(\hbar/2e)Q_m\} = \delta_{lm}$. Between neighboring superconducting layers is an insulating layer, creating the Josephson junctions. There are two types of Josephson junctions in our model, labeled as "1" and "2" or equivalently "weak" and "strong." These are of thickness $d_{1,2}$, and consist of one unit cell of thickness $d = d_1 + d_2$ (Fig. 1). Each junction is characterized by a high-frequency dielectric constant $\epsilon_{1,2}$ and a Josephson critical current $j_{1,2}$. The average dielectric constant for the unit cell is $\epsilon_{\text{av}}^{-1} = (d_1\epsilon_1^{-1} + d_2\epsilon_2^{-1})/d$, and the average capacitance is $C_{\text{av}} = W\epsilon_{\text{av}}/4\pi d$. $W$ is the area of the layers. In addition to the self-charging energy that arises from the nonzero compressibility of each superconducting layer, there are long-range Coulomb interactions among the excess charges $Q_m$ due to lack of screening charges in the insulating layers. The total Hamiltonian is [13, 14]

$$H = -\sum_{m>n} \frac{|x_m - x_n|}{2dC_{\text{av}}} Q_m Q_n + \sum_m \frac{\kappa}{2C_{\text{av}}} Q_m^2 \\ - \sum_m \frac{\hbar j_m^{m+1} W}{e^*} \cos(\theta_{m+1} - \theta_m), \quad (6)$$

where $x_m$ is the coordinate of the $m$th SC layer, with $x_{2m} = md$, and $x_{2m+1} = md + d_1$. The first term represents the Coulomb interactions. The second term is the self-charging energy with a dimensionless compressibility $\kappa = \epsilon_{\text{av}}\mu^2/sd$, with $s$ being the thickness of the superconducting layer. $\mu$ is the Thomas-Fermi screening length in the superconducting layers. The last term is the Josephson energy term. $j_m^{m+1} = j_{1,2}$ is the Josephson critical current between the $m$th and the $(m+1)$th SC layer. The Hamilton equations obtained from Eq. (6) are coupled sine-Gordon equations for $2N$ junctions [32]. Here we show the simplified equations where all the weak/strong junctions are set to be equal, i.e., $\varphi_1 = \varphi_3 = \ldots$ and $\varphi_2 = \varphi_4 = \ldots$,

$$\begin{bmatrix} \ddot{\varphi}_1 \\ \ddot{\varphi}_2 \end{bmatrix} + \gamma \begin{bmatrix} \dot{\varphi}_1 \\ \dot{\varphi}_2 \end{bmatrix} - \frac{4\pi e^* \mu^2 I}{s} \begin{bmatrix} \alpha_1^{-1} \\ \alpha_2^{-1} \end{bmatrix} \\ = \begin{bmatrix} -(1+2\alpha_1)\Omega_1^2 & 2\alpha_2\Omega_2^2 \\ 2\alpha_1\Omega_1^2 & -(1+2\alpha_2)\Omega_2^2 \end{bmatrix} \begin{bmatrix} \varphi_1 \\ \varphi_2 \end{bmatrix}. \quad (7)$$

$\alpha_i$ is the capacitive coupling constant $\alpha_i = \epsilon_i\mu^2/sd_i$, and $\Omega_{1,2}$ is the bare plasma frequency of each junction [see the expression above Eq. (1)]. The parametric driving is included by changing the critical currents $j_i$ as $j_{1,2} \to j_{1,2}[1 \pm A_{1,2}\cos(\omega_{\text{pump}}t)]$; we assume that the driving is alternating along the junctions. The voltage is related to the phase differences by the generalized Josephson relations [14, 24],

$$\left(\frac{\hbar}{e^*}\right) \begin{bmatrix} \dot{\varphi}_1 \\ \dot{\varphi}_2 \end{bmatrix} = \begin{bmatrix} 1+2\alpha_1 & -2\alpha_2 \\ -2\alpha_1 & 1+2\alpha_2 \end{bmatrix} \begin{bmatrix} V_1 \\ V_2 \end{bmatrix}. \quad (8)$$

For the undriven case at $T = 0$, Eqs. (7) and (8) give

$$\sigma(\omega) = \frac{\epsilon_{\text{av}}}{4\pi i} \frac{(\omega^2 + i\gamma\omega - \omega_{\text{Jp1}}^2)(\omega^2 + i\gamma\omega - \omega_{\text{Jp2}}^2)}{\omega(\omega^2 + i\gamma\omega - \omega_t^2)}, \quad (9)$$

where $\omega_{\text{Jp1, Jp2}} \simeq \Omega_{1,2}$ are the longitudinal plasma modes for weak and strong junctions, and $\omega_t \simeq \omega_{\text{Jp2}}$ is the transverse plasma mode [14, 32]. The overall Josephson coupling at $T = 0$, defined as Eq. (5), is $J_0 = e^*\epsilon_{\text{av}}\omega_{\text{Jp1}}^2\omega_{\text{Jp2}}^2/(4\pi\hbar d\omega_t^2)$.

We calculate the effective Josephson coupling for the driven state, similar to the single junction case. Assuming $\omega_{\text{Jp1}}, \Omega_1 \ll \omega_{\text{Jp2}}, \Omega_2$ and $A_i, \gamma \ll 1$, and with a pump frequency near the lower resonance, $\omega_{\text{pump}} \approx \omega_{\text{Jp1}}$, the relative change of the Josephson coupling $\delta J/J_0$ is approximately [32]

$$\delta J/J_0 \simeq \frac{A_1^2\omega_{\text{Jp1}}^2}{2(\omega_{\text{pump}}^2 - \omega_{\text{Jp1}}^2)}(1 + 2\alpha_1 + 2\alpha_2). \quad (10)$$

This indicates that increasing both capacitive couplings further enhances the Josephson coupling. Near the higher resonance $\omega_{\text{pump}} \simeq \omega_{\text{Jp2}}$, we find

$$\delta J/J_0 \simeq \frac{-2\alpha_2^2 A_1^2 + \alpha_1(1+2\alpha_2)A_2^2 + 4\alpha_1\alpha_2 A_1 A_2}{2\alpha_2(\omega_{\text{pump}}^2 - \omega_{\text{Jp2}}^2)}\Omega_1^2. \quad (11)$$

In the pump-probe experiment on YBCO [2], the driving frequency is above $\omega_{\text{Jp2}}$. For that frequency regime, we propose to use a material with a smaller $\alpha_2$ or a larger $\alpha_1$ to exploit the singular behavior of Eq. (11).

Next, we solve the $2N$ coupled sine-Gordon equations with Langevin noise terms numerically to study thermal effects. We focus on the response below the critical temperature since the model assumes condensed Cooper pairs; only the interlayer phase fluctuations are active. We do not address the physics above the critical temperature. For the numerical integration, we employ a Heun scheme with typical time steps of $h = 10^{-5}$, with $N = 50$ unit cells. The other parameters of the model are $\alpha_1 = 3$, $\alpha_2 = 1.5$, $\Omega_1 = 1$, $\Omega_2 = 12.5$, and $\gamma = 0.5$. These are chosen to reproduce the ratio $\omega_{\text{Jp2}}/\omega_{\text{Jp1}} \sim 15.8$ of YBCO with appropriate $\alpha$ values for this compound of around $\sim 3$ [14]. We have $\omega_{\text{Jp1}} = 1.58$ and $\omega_{\text{Jp2}} = 25.1$. We use a small $\gamma < \Omega_i$ to simulate the underdamped regime. To calculate the conductivity $\sigma(\omega_{\mathrm{probe}})$, we add a monochromatic probing current at $\omega_{\mathrm{probe}}$ with $I = 0.1$ [34], and measure the voltage across the junctions. The numerically obtained response functions in equilibrium are in good agreement with the analytical expressions in Eq. (9) [32]. Because of the low-dimensional nature of the system, it does not undergo a true phase transition at nonzero temperature. However, the Josephson coupling decreases rapidly near the crossover temperature $T \sim 1.0$; see Fig. 3(c) [35].

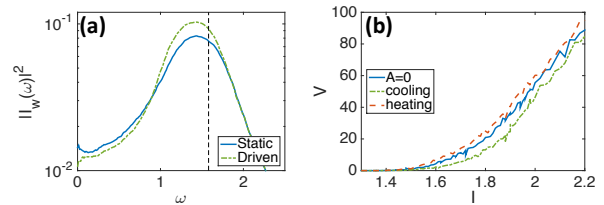

FIG. 4: (a) Power spectrum of the weak junction current at $T = 0.6$ around $\omega_{\mathrm{Jp1}}$; (b) dc current-voltage characteristics.

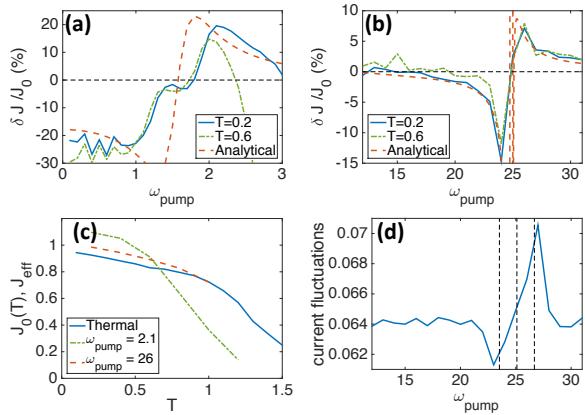

FIG. 3: (a) The relative change of the Josephson coupling $\delta J/J_0(\%)$ near $\omega_{\mathrm{pump}} \simeq \omega_{\mathrm{Jp1}}$ at $A_1 = 0.6$ and $A_2 = 0.3$. (b) The relative change of the Josephson coupling $\delta J/J_0(\%)$ near $\omega_{\mathrm{pump}} \simeq \omega_{\mathrm{Jp2}}$ at $A_1 = 1.0$ and $A_2 = 0.3$. (c) The temperature dependence of $J_0$ and $J_{\mathrm{eff}}$ [normalized by $J_0(0)$] at fixed driving frequencies, $\omega_{\mathrm{pump}} = 2.1$ and $26$. (d) The current fluctuations at the weak junctions, $\langle \sin^2 \varphi_i \rangle_{i \in \mathrm{weak}}$, near $\omega_{\mathrm{pump}} \simeq \omega_{\mathrm{Jp2}}$. The dashed lines are at $\omega_{\mathrm{Jp2}}, \omega_{\mathrm{Jp2}} \pm \omega_{\mathrm{Jp1}}$.

The effective Josephson coupling obtained from the conductivity at $T = 0.2$ and $0.6$ are plotted in Figs. 3(a) and 3(b) near the low and high plasma frequencies. We choose maximal driving amplitudes that are still inside the dynamically stable regions, $(A_1, A_2) = (0.6, 0.3)$ and $(1.0, 0.3)$, respectively. We also plot the analytical solution, including three harmonics, at $T = 0$ for comparison. Similar to the single junction, we observe an enhancement of $J_{\mathrm{eff}}$ just above the two plasma frequencies. The largest relative increases are 20% and 8%, respectively, for these examples, in the steady state. Near the lower resonance, the deviation from the analytical solution is more significant than for the higher resonance, since the higher harmonics become more important in this case. The analytical expression agrees better with the $T = 0.2$ results, and deviates further from the curves for $T = 0.6$ due to thermal effects, especially for the lower resonance. The temperature dependence of $J_{\mathrm{eff}}$ at fixed driving frequencies $\omega_{\mathrm{pump}} = 2.1$ and $26$ is depicted in Fig. 3(c). The enhanced Josephson coupling rapidly decreases as the temperature increases above the lower resonance at $\omega_{\mathrm{pump}} = 2.1$. Near the higher resonance, the enhancement continues up to the crossover temperature of $T_c \simeq 1.0$.

Since the weak junctions dominate the low frequency behavior of the system, we plot the current fluctuations in the weak junctions, $\langle \sin^2 \varphi_i \rangle_{i \in \mathrm{weak}}$, in Fig. 3(d). We denote temporal averages by bars and spatial averages by angle brackets. The pump frequency is near $\omega_{\mathrm{pump}} \sim \omega_{\mathrm{Jp2}}$, close to the experimental conditions of Ref. 2. Near $\omega_{\mathrm{pump}} \sim \omega_{\mathrm{Jp2}} \pm \omega_{\mathrm{Jp1}}$, the current fluctuations increase (decrease) due to parametric heating (cooling) as has been observed in Ref. 6. The enhanced Josephson coupling near $\omega_{\mathrm{pump}} \simeq 26$ occurs when current fluctuations increase due to parametric heating. For an equilibrium system, this would be unexpected, because larger fluctuations are achieved for higher temperatures and lead to a smaller superfluid density and Josephson coupling. In equilibrium, the thermal fluctuations for all frequencies are controlled by a single parameter, the temperature. Here, however, we have created a genuine nonequilibrium state. To demonstrate this, we generate the power spectrum of the weak junction currents, which is the Fourier transform of the temporal current-current correlation function $\overline{I_{\mathrm{w}}(t) I_{\mathrm{w}}(0)}$ where $I_{\mathrm{w}} \equiv \langle \sin \varphi_i \rangle_{i \in \mathrm{weak}}$. This spectrum is shown in Fig. 4 (a) near the parametric heating regime $\omega_{\mathrm{pump}} = 26$. The low frequency part of the power spectrum is related to the conductivity [32], while its sum over all frequencies is equal to the variance $\langle \sin^2 \varphi_i \rangle_{i \in \mathrm{weak}}$. We observe that the low frequency part ($\omega < \omega_{\mathrm{Jp1}}$) has reduced fluctuations, as if the temperature were reduced. This leads to the enhancement of conductivity. However, the total area of the spectrum is increased because the high frequency fluctuations are increased, as if their temperature was increased [36]. We note that the dc current response, which serves as another estimate of Josephson energy [6], is determined by the total power spectrum, and therefore may be different from $J_{\mathrm{eff}}$, which is controlled by the low frequency part. Indeed, we find that the parametric heating (cooling) gives more (fewer) phase slips resulting in larger (smaller) voltages near the critical current, as shown in the dc current-voltage curves in Fig. 4(b). This agrees with the results of Ref. [6].

In conclusion, we have demonstrated a mechanism for light-enhanced superconductivity in which the probe pulse is coupled to a plasma mode by parametric driving. We have first showed this phenomenon for a single, parametrically driven Josephson junction. We have then expanded this analysis to a bilayer system of Josephson junctions which represents the phase and charge dynamics along the $c$ axis of YBCO. Both of these models indeed show an enhanced Josephson coupling when the pump frequency is above the plasma frequencies. We find that smaller damping increases the enhancement, while too little damping induces a parametric instability. We have also proposed material properties that are beneficial for an enhanced Josephson coupling. Finally, we have demonstrated that the resulting driven state is a genuine nonequilibrium state in which enhanced Josephson coupling and increased current fluctuations coexist. We emphasize that the conceptual features of our mechanism provide guidance for a wide range of driven solid state or other many-body systems, in particular for the dynamical control of their low-frequency response functions.

This work was supported by the Deutsche Forschungsgemeinschaft through the SFB 925 and the Hamburg Centre for Ultrafast Imaging, and from the Landesexzellenzinitiative Hamburg, supported by Joachim Herz Stiftung. We thank W. Hu, Y. Laplace, D. Nicoletti, C. Zerbe, R. Höppner, T. Rexin, and B. Zhu for helpful discussions.

---

# Supplemental Material: Theory of enhanced interlayer tunneling in optically driven high $T_c$ superconductors

Jun-ichi Okamoto,[1,2,∗] Andrea Cavalleri,[3,4] and Ludwig Mathey[1,2]

[1]*Zentrum für Optische Quantentechnologien und Institut für Laserphysik,*

*Universität Hamburg, 22761 Hamburg, Germany*

[2]*The Hamburg Centre for Ultrafast Imaging,*

*Luruper Chaussee 149, 22761 Hamburg, Germany*

[3]*Max Planck Institute for the Structure and Dynamics of Matter, 22761 Hamburg, Germany*

[4]*Department of Physics, Clarendon Laboratory,*

*University of Oxford, Oxford OX1 3PU, United Kingdom*

# I. EFFECTS OF HIGHER HARMONICS IN EQ. (2)

In this section, we discuss the effects of higher harmonics in Eq. (2) in the main text. The probing current $I$ oscillates at a frequency $\omega_{\text{probe}}$, and it is coupled to phases $\varphi_n \equiv \varphi(\omega_{\text{probe}} + n\omega_{\text{pump}})$ for an arbitrary integer $n$. To calculate the conductivity and the effective Josephson coupling, we need to know $\varphi_0 \equiv \varphi(\omega_{\text{probe}})$ by solving the set of equations

$$\omega_{\text{Jp}}^{-2} K_n \varphi_n = -\varphi_n - \frac{A}{2}(\varphi_{n-1} + \varphi_{n+1}) + \tilde{I}_{\text{probe}}\delta_{n0}, \tag{1}$$

with $K_n = -(\omega_{\text{probe}} + n\omega_{\text{pump}})^2 - i\gamma(\omega_{\text{probe}} + n\omega_{\text{pump}})$. Approximately we can solve these linear equations by ignoring $\varphi_{m>|n|}$ and $\varphi_{m<-|n|}$ for a chosen integer $n$. Here we use $n = \pm 1, 2$ and 5, and the obtained Josephson coupling as a function of the driving frequency $\omega_{\text{pump}}$ is plotted in Fig. 1 for $A = 0.4$ and $0.9$ with $\gamma = 0.05$. At $A = 0.4$, the basic features around $\omega_{\text{pump}} \sim \omega_{\text{Jp}}$ are well converged already at $n = 1$, while the higher harmonics create structures at lower harmonics $\omega_{\text{pump}} \sim \omega_{\text{Jp}}/2$ for $A = 0.9$.

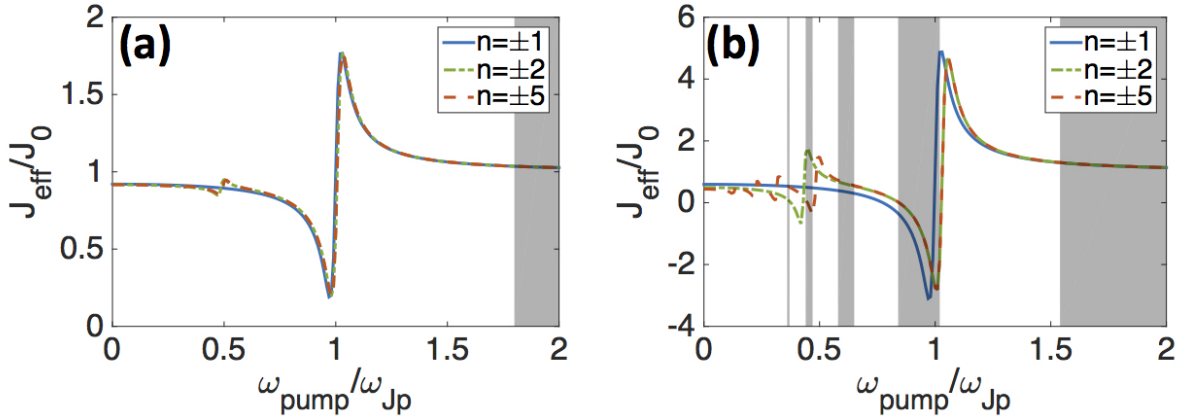

FIG. 1. (a) $J_{\text{eff}}/J_0$ at $A = 0.4$ for cases including up to $n = \pm 1, 2$, and 5. (b) $J_{\text{eff}}/J_0$ at $A = 0.9$. Dynamically unstable regions are indicated by shaded regions.

# II. FLOQUET STABILITY ANALYSIS

Here we briefly explain the Floquet stability analysis[1,2]. According to the Floquet theorem, a first order, linear differential equation

$$\vec{z} = \mathbf{A}(t)\vec{z}, \tag{2}$$

with a periodic coefficient matrix $\mathbf{A}(t) = \mathbf{A}(t + T)$ has a solution in a form

$$z_i(t) = e^{(\ln \lambda_i)t/T} p_i(t) \tag{3}$$

with $\vec{p}(t) = \vec{p}(t + T)$. Here $\lambda_i$ is an eigenvalue of the natural fundamental matrix $\mathbf{X}(T)$ given by the solutions of

$$\dot{\mathbf{X}} = \mathbf{A}(t)\mathbf{X} \tag{4}$$

with initial conditions $X(0) = \mathbf{1}$. If any absolute values of $\lambda$ is bigger than 1, Eq. (3) is diverging, indicating dynamical instability.

In our model, we numerically solved the linearized equation of motion

$$\frac{d}{dt}\begin{bmatrix} \varphi \\ \dot{\varphi} \end{bmatrix} = \begin{bmatrix} \dot{\varphi} \\ -\gamma\dot{\varphi} - \omega_{\text{Jp}}^2 \left[1 + A\cos(\omega_{\text{pump}}t)\right]\varphi \end{bmatrix}, \tag{5}$$

with initial conditions $(\varphi, \dot{\varphi}) = (1, 0)$ and $(0, 1)$ from $t = 0$ to $t = 2\pi/\omega_{\text{pump}}$ to find the eigenvalues $\lambda$ of the natural fundamental matrix.

## III. NONLINEAR EFFECTS ON EQ. (1), AND LOSS FUNCTIONS

We consider the nonlinear effects in Eq. (1) in the main text. Its Fourier transformation leads to

$$(-\omega^2 - i\gamma\omega)\varphi(\omega) = \int \frac{d\omega'}{2\pi} M(\omega - \omega') \left[\sin\varphi\right]_{\omega'} + \tilde{I}(\omega), \tag{6}$$

where $M(\omega)$ is the Fourier transform of $M(t) = -\omega_{\text{Jp}}^2[1 + A\cos(\omega_{\text{pump}}t)]$. To see the nonlinear effect at the lowest order, we approximate $\sin\varphi \simeq \varphi - \varphi^3/3!$, and use the mean-field decompositions as [again, we limit ourselves to $\varphi(\omega_{\text{probe}} \pm \omega_{\text{pump}}) \equiv \varphi_{\pm 1}$ and $\varphi(\omega_{\text{pump}}) \equiv \varphi_0$]

$$\left[\sin\varphi\right]_{\omega_{\text{probe}}} \simeq \varphi_0 \left[1 - \left\langle \frac{1}{2}|\varphi_0|^2 + |\varphi_{-1}|^2 + |\varphi_{-1}|^2 \right\rangle\right], \tag{7}$$

$$\left[\sin\varphi\right]_{\omega_{\text{probe}} \pm \omega_{\text{pump}}} \simeq \varphi_{\pm 1} \left[1 - \left\langle \frac{1}{2}|\varphi_{\pm 1}|^2 + |\varphi_0|^2 + |\varphi_{\mp 1}|^2 \right\rangle\right]. \tag{8}$$

We then have a set of equations, which needs to be solved self-consistently,

$$\begin{bmatrix} K_1 & 0 & 0 \\ 0 & K_0 & 0 \\ 0 & 0 & K_{-1} \end{bmatrix} \begin{bmatrix} \varphi_1 \\ \varphi_0 \\ \varphi_{-1} \end{bmatrix} = -\omega_{\text{Jp}}^2 \begin{bmatrix} 1 & A/2 & 0 \\ A/2 & 1 & A/2 \\ 0 & A/2 & 1 \end{bmatrix} \begin{bmatrix} \tilde{\varphi}_1 \\ \tilde{\varphi}_0 \\ \tilde{\varphi}_{-1} \end{bmatrix} + \begin{bmatrix} 0 \\ \tilde{I} \\ 0 \end{bmatrix}, \tag{9}$$

where $K_n = -(\omega_{\text{probe}} + n\omega_{\text{pump}})^2 - i\gamma(\omega_{\text{probe}} + n\omega_{\text{pump}})$, $\tilde{\varphi}_n = \varphi_n(1 - V_n)$, and $V_n$ is the mean-field term expressed by angle brackets in Eqs. (3) and (4). We start from $V_n = 0$, and solve the above equations to get $\varphi$. Then we use this value to get new value of $V_n$, and then solve the equations again. We repeat the procedures until we get a converged result. We compare the effective Josephson coupling in Fig. 2 for linear and nonlinear models. For a linear model, we find unphysical regions where $J_{\text{eff}}$ becomes negative, which disappear in the nonlinear model. The enhancement of Josephson coupling above $\omega_{\text{Jp}}$ is bigger in the linear model than the nonlinear model. These observations comes from the fact that the diverging behavior $\varphi \sim (\omega_{\text{pump}}^2 - \omega_{\text{Jp}}^2)^{-1}$ is less pronounced in the nonlinear model, since the mean-field terms in Eqs. (3) and (4) reduce the amplitude of $\varphi$'s once they become large.

The loss function is measured similarly by the probing current, and is defined as

$$L(\omega_{\text{probe}}) \equiv -\operatorname{Im}[\omega_{\text{probe}}/4\pi i\sigma(\omega_{\text{probe}})] \propto \operatorname{Im}[\varphi(\omega_{\text{probe}})/\tilde{I}_{\text{probe}}]. \tag{10}$$

In Fig. 3, we compare the loss functions for linear and nonlinear models. $L(\omega_{\text{probe}})$ shows a normal absorption peak around $\omega_{\text{probe}} \simeq \omega_{\text{Jp}}$. We also find a trough/peak around $\omega_{\text{pump}} \simeq \omega_{\text{probe}} \pm \omega_{\text{Jp}}$, since, at these conditions, the parametric driving amplifies $\varphi(\omega_{\text{probe}})$ through the mode $\varphi(\omega_{\text{probe}} - \omega_{\text{pump}})$. The sign of the loss function is determined by the phase difference between $\varphi(\omega_{\text{probe}})$ and $\tilde{I}_{\text{probe}}$. It is usually less than $\pi$ as in a forced harmonic oscillator, and $L(\omega)$ is positive. However, for $\omega_{\text{pump}} \simeq \omega_{\text{probe}} + \omega_{\text{Jp}}$, the phase difference becomes greater than $\pi$ leading to negative values of $L(\omega)$. We note that a negative loss function near $\omega_{\text{pump}} \simeq 2\omega_{\text{Jp}}$ was observed in a LaBaCuO material in Ref. 3. In the linear model, we find an extra resonance around $\omega_{\text{pump}} = -\omega_{\text{probe}} + \omega_{\text{Jp}}$, which disappears for the nonlinear case. Another nonlinear effect is the shift of resonance peak near $\omega_{\text{pump}} \sim \omega_{\text{Jp}}$; as the amplitude $A$ becomes larger, the resonance frequency is pushed to lower frequencies [Fig. 3(d)]. We also find that the minimum of loss function near the dynamical instability is shifted to lower $\omega_{\text{pump}}$ in the nonlinear case as the amplitude gets larger.

## IV. EFFECT OF $\gamma$ FOR EQS. (10) AND (11)

Here we give detailed expressions including the effect of damping $\gamma$ for Eqs. (10) and (11) in the main text. Expanding the lengthy analytical solutions (obtained by considering three harmonics) by $A_{1,2} \sim 0$ and $\gamma \simeq 0$, near the lower resonance, $\omega_{\text{pump}} \simeq \omega_{\text{Jp1}}$, we have

$$\delta J/J_0 \simeq \frac{A_1^2 \omega_{\text{Jp1}}^2}{2(\omega_{\text{pump}}^2 - \omega_{\text{Jp1}}^2)} \left[ (1 + 2\alpha_1 + 2\alpha_2) + \frac{\omega_{\text{Jp1}}^2 (2\alpha_2 + 1)^2}{(\omega_{\text{pump}}^2 - \omega_{\text{Jp1}}^2)^2} \gamma^2 \right] + \mathcal{O}(\omega_{\text{Jp1}}/\omega_{\text{Jp2}}). \quad (11)$$

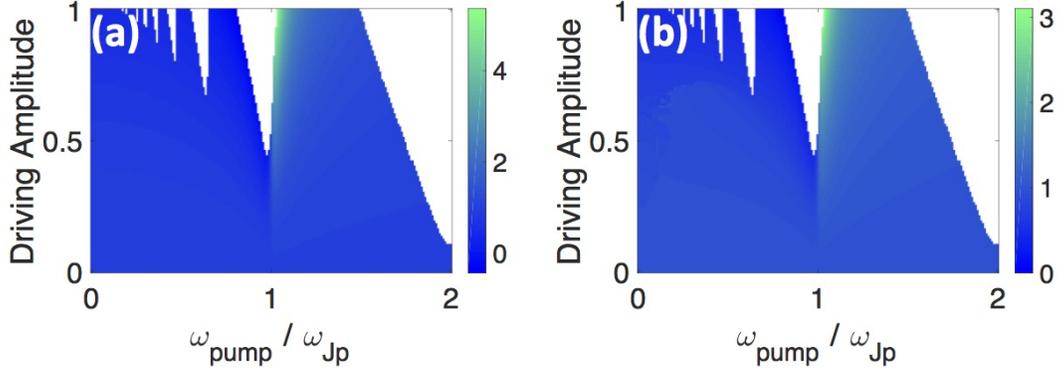

FIG. 2. Superfluid density without nonlinear effects (a) and with nonlinear effects (b). Dynamically unstable regions are excluded.

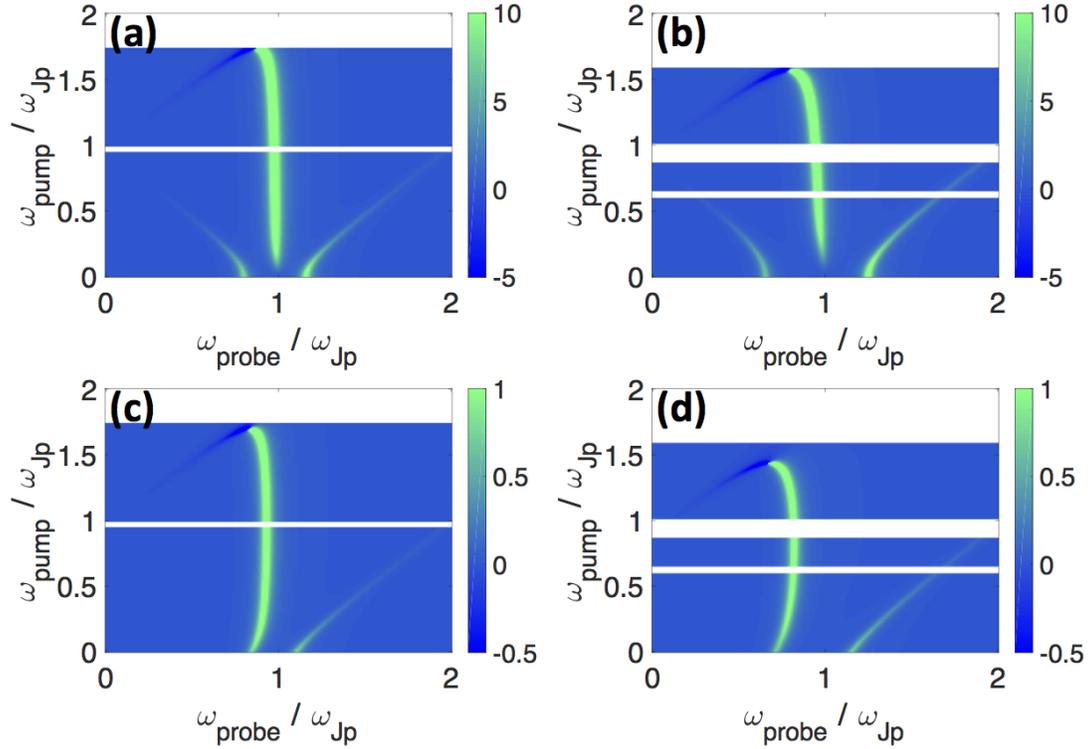

FIG. 3. (a), (b): Loss functions for a linear model at $A = 0.5$ and $A = 0.8$. (c), (d): Loss functions for a nonlinear model at $A = 0.5$ and $A = 0.8$. Dynamically unstable regions are excluded.

This still diverges at $\omega_{\text{pump}} = \omega_{\text{Jp1}}$ since we treat $\gamma$ perturbatively. Near the higher resonance $\omega_{\text{pump}} \simeq \omega_{\text{Jp2}}$, we have

$$
\delta J/J_0 \simeq \frac{-2\alpha_2^2 A_1^2 + \alpha_1(1 + 2\alpha_2)A_2^2 + 4\alpha_1\alpha_2 A_1 A_2}{2\alpha_2(\omega_{\text{pump}}^2 - \omega_{\text{Jp2}}^2)}\Omega_1^2
$$
$$
- \frac{\alpha_1\left(2\alpha_2 + 1\right)A_2^2 + 4\alpha_1\alpha_2 A_1 A_2 + 4\alpha_2^2\left[\alpha_1\left(1 - 2\alpha_2\right) - 2\alpha_2^2\right]A_1^2}{2\alpha_2(\omega_{\text{pump}}^2 - \omega_{\text{Jp2}}^2)^3}\Omega_1^2\Omega_2^2\gamma^2 + \mathcal{O}(\omega_{\text{Jp1}}/\omega_{\text{Jp2}}).
$$

$$(12)$$

## V. COUPLED SINE-GORDON EQUATIONS WITH LANGEVIN NOISES

Here we show the details of the coupled sine-Gordon equation used for our simulations. The starting Hamilton equations, based on the Hamiltonian in Eq. (6) in the main text, are[12]

$$
\left(\frac{\hbar}{e^*}\right)\dot\theta_m = \sum_n \frac{1}{2C_{\text{av}}}\left(-\frac{|x_m - x_n|}{d} + 2\kappa\delta_{mn}\right)Q_n,
$$
$$
\left(\frac{\hbar}{e^*}\right)\dot Q_m = \frac{\hbar W}{e^*}\left[j_m^{m+1}\sin\varphi_{m+1,m} - j_{m-1}^m\sin\varphi_{m,m-1}\right],
$$

$$(13)$$

where we introduced a phase difference between the $m$th and $(m+1)$th SC layers as $\varphi_{m+1,m} = \theta_{m+1} - \theta_m$. Eliminating $\{Q_m\}$ leads to the coupled sine-Gordon equations derived in Refs. 4 and 5

$$
\ddot{\vec{\varphi}} \equiv \begin{bmatrix} \ddot\varphi_{10} \\ \ddot\varphi_{21} \\ \ddot\varphi_{32} \\ \vdots \end{bmatrix} = \begin{bmatrix} -(1 + 2\alpha_1)\Omega_1^2 & \alpha_2\Omega_2^2 & & \\ \alpha_1\Omega_1^2 & -(1 + 2\alpha_2)\Omega_2^2 & \alpha_1\Omega_1^2 & \\ & \alpha_2\Omega_2^2 & -(1 + 2\alpha_1)\Omega_1^2 & \alpha_2\Omega_2^2 \\ & & & \ddots \end{bmatrix} \begin{bmatrix} \sin\varphi_{10} \\ \sin\varphi_{21} \\ \sin\varphi_{32} \\ \vdots \end{bmatrix} \equiv \mathbf{M}\vec{J_s}
$$

$$(14)$$

where $\alpha_i$ is the capacitive coupling constant

$$
\alpha_i = \epsilon_i\mu^2/sd_i,
$$

$$(15)$$

and $\Omega_i$ is the bare plasma frequency of a junction,

$$
\Omega_i = \sqrt{\frac{4\pi e^* d_i j_i}{\hbar\epsilon_i}}.
$$

$$(16)$$

We further add the damping term $\gamma$, Langevin thermal noises $\vec{\xi}$, a total external current $I(t)$,

$$\ddot{\vec{\varphi}} + \gamma\dot{\vec{\varphi}} = \mathbf{M}\vec{J_s} + \vec{I_0} + \vec{\xi}, \tag{17}$$

where $\vec{I_0} = 4\pi e^*\mu^2 I/s\hbar(\alpha_1^{-1}, \alpha_2^{-1}, \alpha_1^{-1}, \dots)$. The Langevin noises are correlated as $\langle \xi_i(t)\xi_j(t')\rangle = \delta(t-t')2\gamma c^2 k_B T B_{ij}$ where $B_{ij}$ is the $(i, j)$ component of a matrix $\mathbf{B}$,

$$\mathbf{B} = \frac{16\pi^3\mu^2}{\phi_0^2 Ws}\begin{bmatrix} 2+\alpha_1^{-1} & -1 & & & \\ -1 & 2+\alpha_2^{-1} & -1 & & \\ & -1 & 2+\alpha_1^{-1} & -1 & \\ & & & \ddots \end{bmatrix}, \tag{18}$$

$k_B$ is the Boltzman constant and $\phi_0 = hc/e^*$ is the flux quantum. For the simulations, we normalize the temperature by $\hbar j_1 W/\alpha_1\Omega_1^2 e^*$.

Response functions can be calculated using an external current $I(t) = I_0\cos(\omega_{\text{pump}}t)$, and the voltage response $V(t)$ as discussed in Ref. 6. The voltage is related to the phase differences by generalized Josephson relations:

$$\left(\frac{\hbar}{e^*}\right)\dot{\vec{\varphi}} = \mathbf{\Lambda}\vec{V} \tag{19}$$

with

$$\mathbf{\Lambda} = \begin{bmatrix} 1+2\alpha_1 & -\alpha_2 & & & \\ -\alpha_1 & 1+2\alpha_2 & -\alpha_1 & & \\ & -\alpha_2 & 1+2\alpha_1 & -\alpha_2 & \\ & & & \ddots \end{bmatrix}. \tag{20}$$

The average electric field for the whole junctions is $E_{\text{av}}(t) = V(t)/(dN)$. The numerically obtained response functions are plotted in Fig. 4. For an unperturbed case, the conductivity of a linear model can be calculated analytically at $T = 0$,

$$\sigma(\omega) = \frac{\epsilon_{\text{av}}}{4\pi i}\frac{\left(\omega^2 + i\gamma\omega - \omega_{\text{Jp1}}^2\right)\left(\omega^2 + i\gamma\omega - \omega_{\text{Jp2}}^2\right)}{\omega\left(\omega^2 + i\gamma\omega - \omega_t^2\right)}, \tag{21}$$

where $\omega_{\text{Jp1, Jp2}}$ are two longitudinal plasma modes

$$\omega_{\text{Jp1, Jp2}}^2 = \left(\frac{1}{2}+\alpha_1\right)\Omega_1^2 + \left(\frac{1}{2}+\alpha_2\right)\Omega_2^2$$
$$\mp\sqrt{\left[\left(\frac{1}{2}+\alpha_1\right)\Omega_1^2 - \left(\frac{1}{2}+\alpha_2\right)\Omega_2^2\right]^2 + 4\alpha_1\alpha_2\Omega_1^2\Omega_2^2}, \tag{22}$$

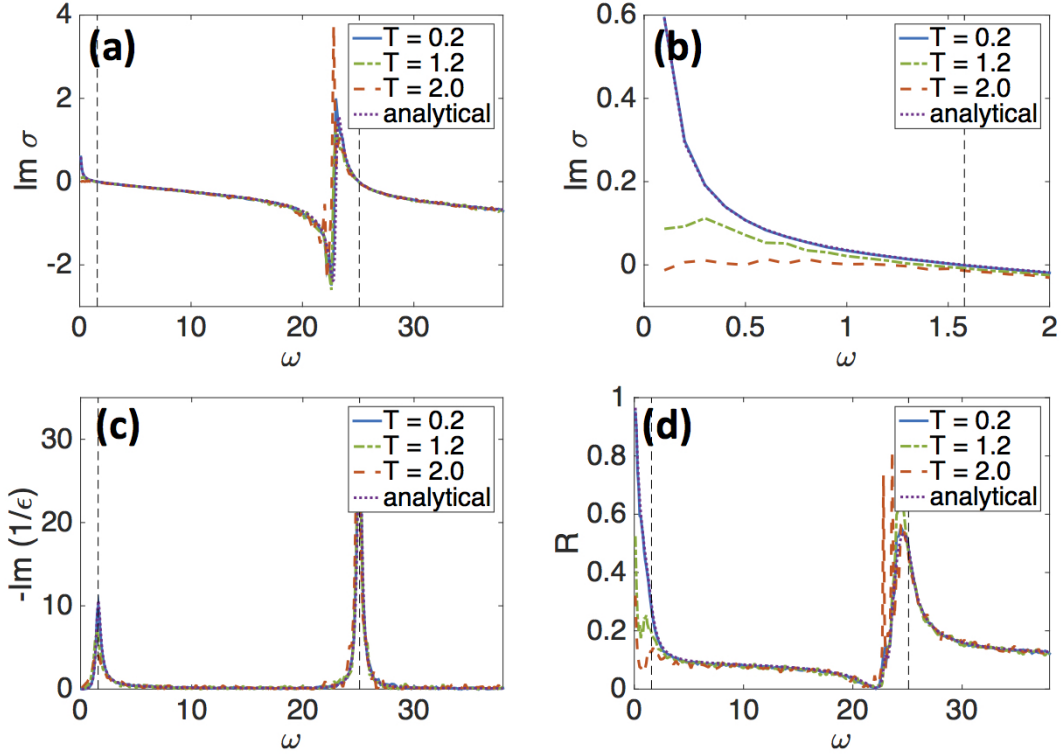

FIG. 4. (a) Imaginary part of conductivity, (b) A closer look of (a) near $\omega_{\mathrm{Jp1}} = 1.58$, (c) loss function, and (d) reflectivity. $\epsilon_{\mathrm{av}} \approx 0.25$ in our unit.

and $\omega_t$ is the transverse plasma mode,

$$\omega_t^2 = \frac{1 + 2\alpha_1 + 2\alpha_2}{\alpha_1 + \alpha_2} \left( \alpha_1 \Omega_1^2 + \alpha_2 \Omega_2^2 \right). \tag{23}$$

Fig. 4 (a) and (b) reproduce the basic features of this analytical solution such as the diverging behavior near $\omega_t$, and $1/\omega$ divergence as $\omega \ll \omega_{\mathrm{Jp1}}$. As the temperature increases, the latter divergence, the signature of superconducting states, disappears around $T = 1.0$. Fig. 4 (c) shows the loss function, and the two peaks correspond to the absorption peaks of two plasma frequencies $\omega_{\mathrm{Jp1, Jp2}}$. Fig. 4 (d) is the reflectivity. While BCS type superconductors show perfect reflection $R \sim 1$ below the gap energy, in a layered compound $R$ gets nearly 1 only at very low frequencies since the screening is not perfect.

## VI.   POWER SPECTRUM FOR A LINEARIZED MODEL: GREEN'S FUNCTION METHOD

Here we outline an approach to obtain power spectrum for a linearized model; Green's function method[1,7,8]. We consider $2N$ coupled parametrically driven Brownian oscillators that obeys the linearized equations of Eq. (17).

$$\ddot{\vec{\varphi}} + \gamma \dot{\vec{\varphi}} - \mathbf{M}(t, \phi)\vec{\varphi} = \vec{\xi},$$   (24)

where $\langle \xi_i(t)\xi_j(s)\rangle = 2\gamma c^2 k_B T B_{ij}\delta(t-s)$, and the driving depends on the initial phase $\phi$. Introducing a new variable $\vec{\varphi} = \vec{y}e^{-\frac{\gamma}{2}t}$, we get

$$\ddot{\vec{y}} - \mathbf{M}(t, \phi)\vec{y} - \frac{\gamma^2}{4}\mathbf{1}_n\vec{y} = \vec{\xi}e^{\frac{\gamma}{2}t} \equiv \vec{\eta}.$$   (25)

Now we make this second order differential equation into a first order equation by using $\vec{z} = (\vec{y}, \dot{\vec{y}})^t$. The equation of motion is found to be

$$\dot{\vec{z}} = \begin{bmatrix} 0 & \mathbf{1}_n \\ \mathbf{M}(t, \phi) + \mathbf{1}_n\frac{\gamma^2}{4} & 0 \end{bmatrix}\vec{z} + \begin{bmatrix} 0 \\ \vec{\eta} \end{bmatrix} \equiv \mathbf{A}(t, \phi)\vec{z} + \vec{f}(t).$$   (26)

The natural fundamental matrix $\mathbf{\Phi}(t, \phi)$ is numerically obtained by solving the equation with the initial condition $\mathbf{\Phi}(0) = \mathbf{1}_{2n}$ without the inhomogeneous term, i.e., $\dot{\mathbf{\Phi}} = \mathbf{A}(t, \phi)\mathbf{\Phi}$. The Green's matrix is defined as

$$\mathbf{G}(t, s, \phi) = \mathbf{\Phi}(t, \phi)\mathbf{\Phi}^{-1}(s, \phi).$$   (27)

Then the solution is given by[1]

$$\vec{z}(t, \phi) = \mathbf{\Phi}(t, \phi)\vec{z}(0) + \int_0^t \mathbf{G}(t, s, \phi)\vec{f}(s)ds$$   (28)

Going back to the original basis $\vec{\varphi}$, we need to be careful that $\vec{y}(0) = \vec{\varphi}(0) + \frac{\gamma}{2}\vec{\varphi}(0)$.

We focus on the homogeneous case, $\varphi_1 = \varphi_3 = \varphi_5 = \cdots$ and $\varphi_2 = \varphi_4 = \varphi_6 = \cdots$. The correlation function of the 1st junction is (for $t > t'$)

$$\langle \varphi_1(t, \phi)\varphi_1(t', \phi)\rangle = \sum_{i', j'=1}^4 \Big[\mathbf{\Phi}_{1i'}(t, \phi)z_{i'}(0)\mathbf{\Phi}_{1j'}(t', \phi)z_{j'}(0)$$
$$+ 2\gamma k_B T \int_0^{t'} \mathbf{G}_{1i'}(t, s, \phi)\mathbf{G}_{1j'}(t', s, \phi)\tilde{\mathbf{B}}_{i'j'}e^{\gamma s}ds\Big]e^{-\frac{\gamma}{2}(t+t')},   (29)$$

where $\vec{z}(0) = (\varphi_1(0), \varphi_2(0), \dot{\varphi}_1(0) + \frac{\gamma}{2}\varphi_1(0), \dot{\varphi}_2(0) + \frac{\gamma}{2}\varphi_2(0))$ and $\langle f_i(s)f_j(s')\rangle = 2\gamma c^2 k_B T \tilde{B}_{ij}\delta(s - s')e^{\gamma s}$ with

$$\tilde{\mathbf{B}} = \begin{bmatrix} 0 & 0 \\ 0 & \mathbf{B} \end{bmatrix}. \qquad (30)$$

The upper limit of the integral is $t'$ since we now consider $t > t'$. In the long-time limit, the steady state is expected to be independent of the initial condition, so we will focus on the second term. The time translation invariance will be recovered after averaging over the phase $\phi$. The power spectrum at $\omega = 26$, $T = 0.6$, $A_1 = 1.0$, and $A_2 = 0.3$ is given in Fig. 5. We see that the spectral weights are reduced for low frequencies, while the total weights, i.e., the sum of weights over all frequencies, are increased. This basically agrees with the power spectrum obtained from Langevin simulations in the main text.

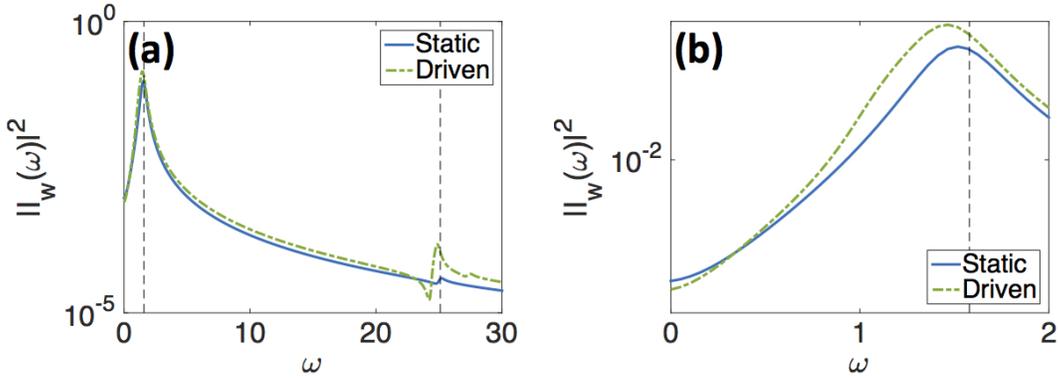

FIG. 5. Power spectrum of static and driven ($\omega_{\text{pump}} = 26.0$) cases at $T = 0.6$ with $A_1 = 1.0$ and $A_2 = 0.3$. (a) a wide view (b) a closer view near $\omega_{\text{Jp1}} = 1.58$.

## VII. LINEAR RESPONSE THEORY

We summarize the linear response theory for bilayer Josephson junctions that obey the stochastic equations of motion in Eq. (17). The corresponding Fokker-Plank equation for a probability density $p(\vec{\varphi}, \dot{\vec{\varphi}}, t)$ is

$$\dot{p} = (\mathcal{L}_0 + \mathcal{L}_1)p, \qquad (31)$$

where $\mathcal{L}_0$ is the unperturbed part

$$
\begin{aligned}
\mathcal{L}_0 = \gamma c^2 k_B T \sum_{ij} & \left( B_{ij} \frac{\partial^2}{\partial \dot\varphi_i \partial \dot\varphi_j} \right) \\
+ \sum_i & \left[ -\dot\varphi_i \frac{\partial}{\partial \varphi_i} + \frac{\partial}{\partial \dot\varphi_i} \left( \gamma \dot\varphi_i + \frac{c^2 \hbar W}{e^*} \sum_j B_{ij} j_j \sin \varphi_j \right) \right],
\end{aligned}
\tag{32}
$$

and $\mathcal{L}_1$ is the perturbation by the probing current

$$
\mathcal{L}_1(t) = - \sum_i I_{0,i}(t) \frac{\partial}{\partial \dot\varphi_i}.
\tag{33}
$$

The deviation of a phase velocity from the equilibrium distributions is related to correlation functions[9]

$$
\begin{aligned}
\delta \langle \dot\varphi_i \rangle (t) &= \sum_j \int_{-\infty}^{\infty} ds R^{ij}(t-s) I_{0,j}(s), \\
R^{ij}(t) &= \frac{1}{c^2 k_B T} \sum_k B_{jk}^{-1} \langle \dot\varphi_i(t) \dot\varphi_k(0) \rangle,
\end{aligned}
\tag{34}
$$

or equivalently $\delta \langle \dot\varphi_i \rangle (\omega) = \sum_j R^{ij}(\omega) I_{0,j}(\omega)$. Defining the velocity susceptibility as $\chi_{ij}(t) = \langle \dot\varphi_i(t) \dot\varphi_j(0) \rangle$, the total voltage difference is

$$
\vec{V}(\omega) = \frac{\hbar}{e^* c^2 k_B T} \mathbf{\Lambda}^{-1} \chi(\omega) \mathbf{B}^{-1} \vec{I_0}(\omega).
\tag{35}
$$

Now for the sake of simplicity we consider spatially homogeneous case. The largest contribution to the voltage is from the weak junctions, so we can approximate the total voltage across the two junctions is

$$
V(\omega) \simeq \frac{\hbar^2}{e^{*2} W k_B T} \chi_{11}(\omega) I(\omega).
\tag{36}
$$

Since the velocity susceptibility is connected to the coordinate susceptibility by the simple time derivative[9], this formula indicates that the lower power spectrum of the current fluctuations at low frequencies leads to a larger conductivity.

———————————

*  ojunichi@physnet.uni-hamburg.de